\documentclass{elsarticle} 
\usepackage{amssymb}
\usepackage[T1]{fontenc}
\usepackage{mathtools}
\usepackage{xfrac}
\usepackage{lmodern}
\usepackage{braket}
\usepackage{physics}
\usepackage[inline]{enumitem}
\usepackage{xcolor}
\usepackage{amsthm}
\usepackage[noabbrev,capitalize]{cleveref}
\usepackage{changepage}
\usepackage{calc}
\usepackage{relsize}

\newtheorem{theorem}{Theorem}
\newtheorem{lemma}[theorem]{Lemma}
\newtheorem{corollary}[theorem]{Corollary}
\newdefinition{definition}{Definition}
\newdefinition{example}{Example}
\crefname{subsection}{subsection}{subsections}
\Crefname{subsection}{Subsection}{Subsections}

\newenvironment{turing}[2]
 {\begin{enumerate}[labelsep=0pt,align=left,parsep=0pt,leftmargin=0pt]
  \item[$#1={}$]``\ignorespaces#2
  \begin{enumerate}[
    nosep,
    labelsep=.5em,
    leftmargin=\widthof{$#1={}$}+1.5em,
    labelwidth=1.5em,
    label=\bfseries\arabic{*}.,
    ref=\arabic{*}
  ]}
 {\unskip''\end{enumerate}\end{enumerate}}

\newcommand{\bitem}[1]{\item\begin{adjustwidth}{1em}{0pt}\ignorespaces#1\end{adjustwidth}}
\newcommand{\bbitem}[1]{\item\begin{adjustwidth}{2em}{0pt}\ignorespaces#1\end{adjustwidth}}

\makeatletter
\newcommand{\subalign}[1]{%
  \vcenter{%
    \Let@ \restore@math@cr \default@tag
    \baselineskip\fontdimen10 \scriptfont\tw@
    \advance\baselineskip\fontdimen12 \scriptfont\tw@
    \lineskip\thr@@\fontdimen8 \scriptfont\thr@@
    \lineskiplimit\lineskip
    \ialign{\hfil$\m@th\scriptstyle##$&$\m@th\scriptstyle{}##$\hfil\crcr
      #1\crcr
    }%
  }%
}
\makeatother

\DeclarePairedDelimiterX\PH[1][]{
   
   #1
}

\usepackage[normalem]{ulem}

\usepackage{tikz}
\usetikzlibrary{automata, positioning, decorations.pathmorphing, calc, quotes}

\tikzset{
	every edge/.append style={
		every node/.append style={
			execute at begin node=$,
			execute at end node=$
		}
	},
	every state/.append style={
		execute at begin node=$,
		execute at end node=$
	},
	>=stealth,
	node distance = .7cm and 3.5cm,
	on grid,
	auto,
	initial text = 
}

\DeclarePairedDelimiter{\ceil}{\lceil}{\rceil}

\renewcommand{\P}[1]{\mathcal{P}\paren*{#1}}

\DeclarePairedDelimiter\paren\lparen\rparen
\DeclarePairedDelimiter\aparen\langle\rangle

\newcommand{\DP}[1]{\ensuremath{\mathit{#1}}}
\newcommand{\DECIDER}[1]{\ensuremath{\DP{HALTING}_{#1}}}

\newcommand{\CC}[1]{\ensuremath{\operatorname{#1}}}

\newcommand{\eL}{\mathcal{L}}
\renewcommand{\L}[1]{\eL\paren*{#1}}

\newcommand{\OH}[1]{\ensuremath{\mathit{O}\paren*{#1}}}

\newcommand{\THETA}[1]{\ensuremath{\Theta\paren*{#1}}}
\newcommand{\THETAp}[2]{\ensuremath{\Theta\paren[#1]{#2}}}
\newcommand{\NSPACE}[1]{\ensuremath{\CC{NSPACE}\paren*{#1}}}
\newcommand{\NTISP}[2]{\ensuremath{\CC{NTISP}\paren*{#1, #2}}}

\newcommand{\NL}{\CC{NL}}

\newcommand{\NP}{\CC{NP}}
\newcommand{\TNFA}[1]{\ensuremath{\L{\tnfa{#1}}}}
\newcommand{\TNFAT}[2]{\ensuremath{\L{\tnfa{#1}, #2}}}
\newcommand{\TNFAL}[1]{\TNFAT{#1}{\timel{linear}}}

\newcommand{\TNFAS}{\ensuremath{\bigcup_{k > 0} \TNFA{k}}}
\newcommand{\TNFAST}[1]{\TNFAT{*}{#1}}
\newcommand{\TNFASL}{\TNFAST{\timel{linear}}}

\newcommand{\RR}[1]{\ensuremath{\mathsf{#1}}}
\newcommand{\timel}[1]{\RR{#1\text{-}time}}

\newcommand{\IP}[1][]{\ensuremath{\CC{{#1}1IP}}}
\newcommand{\IPerr}[2][]{\ensuremath{\IP[#1]_{#2}}}
\newcommand{\IPstar}[1][]{\ensuremath{\IP[#1]_{*}}}
\newcommand{\IPl}[2][]{\ensuremath{\IP[#1]\paren*{#2}}}
\newcommand{\IPerrl}[3][]{\ensuremath{\IPerr[#1]{#2}\paren*{#3}}}
\newcommand{\IPstarl}[2][]{\ensuremath{\IPstar[#1]\paren*{#2}}}

\newcommand{\lcons}{\RR{cons}}
\newcommand{\llog}{\RR{log}}
\newcommand{\lpoly}{\RR{poly}}
\newcommand{\lexp}{\RR{exp}}

\newcommand{\MM}[1]{\textsf{#1}}
\newcommand{\TM}{\MM{TM}}
\newcommand{\NTM}{\MM{NTM}}
\newcommand{\PTM}{\MM{PTM}}
\newcommand{\IPS}{\MM{IPS}}

\newcommand{\onfa}{\MM{1nfa}}
\newcommand{\odfa}{\MM{1dfa}}

\newcommand{\afa}{\MM{2afa}}
\newcommand{\nfa}{\MM{2nfa}}
\newcommand{\dfa}{\MM{2dfa}}
\newcommand{\tnfa}[1]{\MM{\nfa\ensuremath{\paren*{#1}}}}

\newcommand{\tdfa}[1]{\MM{\dfa\ensuremath{\paren*{#1}}}}
\newcommand{\pfa}{\MM{2pfa}}
\newcommand\optk[1]{\ensuremath{k_{#1}}}
\newcommand{\N}{\mathbb{N}}

\newcommand\anbn{\ensuremath{\texttt{EQ}}}
\newcommand\nanb{\ensuremath{\texttt{MIXEDEQ}}}
\newcommand\PAL{\ensuremath{\texttt{PAL}}}
\newcommand\CERT{\ensuremath{\texttt{CERT}}}

\newcommand\PATH{\ensuremath{\texttt{PATH}}}
\newcommand\TA{\ensuremath{x}}
\newcommand\bin{\ensuremath{\Set{0, 1}}}

\newcommand{\oneto}[1]{\ensuremath{\Set{1, \dotsc, #1}}}

\newcommand\closedintrange[2]{\Set{#1, \dotsc, #2}}

\newcommand\prmark[1]{\ensuremath{#1_r}}
\newcommand\detmark[1]{\ensuremath{#1_d}}

\newcommand{\qacc}{\ensuremath{q_{\text{acc}}}}
\newcommand{\qrej}{\ensuremath{q_{\text{rej}}}}
\newcommand{\lend}{\ensuremath{\rhd}}
\newcommand{\rend}{\ensuremath{\lhd}}
\newcommand{\midm}{\ensuremath{\bowtie}}

\newcommand{\tblank}{\texttt{\textvisiblespace}}

\newcommand{\tsh}{\texttt{\#}}
\newcommand{\manbn}[1][]{\ensuremath{M_{\anbn_{#1}}}}
\newcommand\movel{\ensuremath{-1}}
\newcommand\movep{\ensuremath{0}}
\newcommand\mover{\ensuremath{+1}}

\newcommand{\marked}[2][0pt]{\smash{\overset{\raisebox{-12pt}{\hspace{#1}\tiny$\blacktriangledown$}}{#2}}}
\newcommand{\tmext}[1]{{#1}_\diamond}
\newcommand{\enc}[1]{\aparen{#1}}

\newcommand{\Accept}{\textit{Accept}}
\newcommand{\Reject}{\textit{Reject}}
\newcommand{\accept}{\textit{accept}}
\newcommand{\reject}{\textit{reject}}

\newcommand{\windable}{risky}

\newcommand{\kulyutmaz}{safe}
\newcommand{\Kulyutmaz}{Safe}

\newcommand\counters{counters}
\newcommand\Counters{Counters}
\newcommand\CVAR{\ensuremath{\kappa}}
\newcommand\cache{cache}

\newcommand\caches{caches}
\newcommand\Caches{Caches}
\newcommand\CACHEVAR{\ensuremath{\eta}}

\newcommand{\recache}{re-cache}
\newcommand{\recached}{re-cached}
\newcommand{\recaches}{re-caches}
\newcommand{\Recache}{Re-cache}
\newcommand{\Recaches}{Re-caches}
\newcommand{\recaching}{re-caching}

\newcommand{\ith}[2][th]{\ensuremath{#2}\textsuperscript{#1}}
\newcommand{\step}[1]{stage~\ref{#1}}
\newcommand{\Step}[1]{Stage~\ref{#1}}

\newcommand{\Steps}[2]{Stages~\ref{#1} and \ref{#2}}

\newcommand{\textiff}{if and only if}

\newcommand\impit[2]{\ref{#1}${}\implies{}$\ref{#2}}

\newcommand{\err}{\ensuremath{\mathlarger{\mathlarger{\varepsilon}}}}
\newcommand{\errf}[1]{\ensuremath{\err_{#1}}}

\newcommand\windmark[1]{\ensuremath{#1_{\mathchoice{}{}{\scriptscriptstyle}{}R}}}
\newcommand\relmark[1]{\ensuremath{#1_{\mathchoice{}{}{\scriptscriptstyle}{}S}}}
\newcommand{\kw}{\windmark{k}}
\newcommand{\kr}{\relmark{k}}

\newcommand\opw{\windmark{P}}

\newcommand\SYW{\ensuremath{\mu_1}}
\newcommand\SYS{\ensuremath{\mu_2}}
\newcommand\GB{\ensuremath{\mu_3}}

\newcommand\SYWf[1]{\ensuremath{\SYW\paren{#1}}}
\newcommand\SYSf[1]{\ensuremath{\SYS\paren{#1}}}
\newcommand\GBf[1]{\ensuremath{\GB\paren{#1}}}

\setlist{parsep=0.1em, itemsep=0.3em}
\newcommand\POS\rho

\newcommand\C{\ensuremath{m}} % For how many iterations to simulate
\newcommand\B{\ensuremath{r}} % #coins to flip for head-choice
\newcommand\SB{\ensuremath{s}} % #coins to flip for head-type-choice
\newcommand\selfun{\nu} % LUT for head-choice
\newcommand\selfunw{\ensuremath{\windmark\selfun}} % LUT for windable head-choice
\newcommand\selfunr{\ensuremath{\relmark\selfun}} % LUT for kulyutmaz head-choice
\newcommand\sffw[1]{\ensuremath{\selfunw\paren*{#1}}} % LUT for head-type-choice
\newcommand\sffr[1]{\ensuremath{\selfunr\paren*{#1}}} % LUT for head-type-choice

\newlist{todolist}{itemize}{2}
\setlist[todolist]{label=$\square$}
\usepackage{pifont}
\begin{document}
% \fontsize{10pt}{13.5pt}
% \selectfont

\begin{frontmatter}

\title{Constant-Space, Constant-Randomness Verifiers with Arbitrarily Small Error\tnoteref{lataref}}
\tnotetext[lataref]{This paper is a substantially improved version of~\cite{gezer}.}
% \titlerunning{Const.-Space, Const.-Randomness Verifiers with Arbitrarily Small Error}

\author{M. Utkan Gezer\corref{cor1}}
\ead{utkan.gezer@boun.edu.tr}

\author{A. C. Cem Say}
\ead{say@boun.edu.tr}

\cortext[cor1]{Corresponding author}
\address{Department of Computer Engineering, Bo\u{g}azi\c{c}i University, Bebek 34342, \.{I}stanbul, Turkey}

\begin{abstract}
     We study the capabilities of probabilistic finite-state  machines that act as verifiers  for certificates of language membership for input strings, in the regime where the verifiers are restricted to toss some fixed nonzero number of coins regardless of the input size. Say and Yakary\i{}lmaz showed that the class of languages that could be verified by these machines within an error bound strictly less than $\sfrac12$ is precisely \NL, but their construction yields verifiers with error bounds that are very close to $\sfrac12$ for most languages in that class when the definition of ``error'' is strengthened to include looping forever without giving a response. 
     We characterize a subset of \NL{} for which verification with arbitrarily low error is possible by these extremely weak machines. It turns out that, for any  $\err>0$,  one can construct a constant-coin, constant-space verifier operating within error $\err$ for every language that is recognizable by a linear-time multi-head nondeterministic finite automaton (\tnfa{k}). We discuss why it is difficult to generalize this method to all of \NL, and give a reasonably tight way to relate the power of linear-time \tnfa{k}'s to simultaneous time-space complexity classes defined in terms of Turing machines. 
\end{abstract}


\end{frontmatter}

\section{Introduction}
The classification of languages in terms of the resources required for verifying proofs (``certificates'') of membership in them is a major concern of computational complexity theory. In this context, important tradeoffs among different types of resources such as time, space, and randomness have been demonstrated: The power of deterministic polynomial-time, polynomial-space bounded verifiers characterized by the class \NP{} has, for instance, been shown to be identical to that of probabilistic bounded-error polynomial-time logarithmic-space verifiers that toss only logarithmically many coins in terms of the input size~\cite{cl95}.  

The study of finite-state probabilistic verifiers started in the late 1980's. Condon and Lipton~\cite{condonlipton} showed that, even under this severe space restriction, one can verify membership in any Turing-recognizable language if one is not required to halt with high probability on rejected inputs. Dwork and Stockmeyer~\cite{dworkstock} showed that interactive proof systems with constant-space verifiers outperform ``stand-alone'' finite-state recognizers when required to halt with high probability as well. The area has grown to have a rich literature where scenarios with multiple provers and quantum verifiers have also been considered. The study of interactive proof systems with quantum finite automata, which was initiated by Nishimura and Yamakami~\cite{nishimura2009,nishimura2015}, continued with the consideration of more powerful quantum models by Yakary\i{}lmaz~\cite{yakaryilmazqAM} and Zheng et al.~\cite{zheng}. The power of finite-state verifiers that are faced with two opposing provers were studied by Feige and Shamir~\cite{feige} and Demirci et al.~\cite{debate} in the classical setup, and by Yakary\i{}lmaz et al.~\cite{transparent} in the quantum setup.

Recently, Say and Yakary\i{}lmaz initiated the study of the power of classical finite-state verifiers that are restricted to toss some fixed, nonzero number of coins regardless of the input size, and proved~\cite{sayyakaryilmaz} that the class of languages which have certificates that could be verified by these machines within an error bound strictly less than $\sfrac12$ is precisely \NL, i.e.\ languages with deterministic logarithmic-space verifiers.

% The construction given in \cite{sayyakaryilmaz} could exhibit a constant-randomness verifier operating within error $\err$ for some $\err < \sfrac12$ for any language in \NL, however, it did not provide a method for reducing this error to more desirable smaller values. Indeed, for many languages in \NL, the constructed verifier's error bound is uncomfortably close to $\sfrac12$, raising the question of whether the class of languages for which it is possible to obtain verifiers with arbitrarily small positive error bounds is a proper subset of \NL{} or not.
The construction given in~\cite{sayyakaryilmaz} could exhibit a constant-randomness verifier operating within error $\err$ for some $\err < \sfrac12$ for any language in \NL, however, it provided a method for reducing this error to  more desirable smaller values only in the ``weak'' regime where looping forever without a response is not considered to be an error.  Indeed, when the error definition is strengthened to include this behavior, for many languages in \NL, the constructed verifier's error bound is uncomfortably close to $\sfrac12$, raising the question of whether the class of languages for which it is possible to obtain verifiers with arbitrarily small positive error bounds is a proper subset of \NL{} or not.

In this paper, we characterize a subset of \NL{} for which verification with   arbitrarily low error  is possible by these extremely weak machines. It turns out that for any $\err>0$, one can construct a constant-coin, constant-space verifier operating within error $\err$ for every language that is recognizable by a linear-time multi-head finite automaton (\tnfa{k}). We discuss why it is difficult to generalize this method to all of \NL{} and give a reasonably tight way to relate the power of linear-time \tnfa{k}'s to simultaneous time-space complexity classes defined in terms of Turing machines. We conclude with a list of open questions.

\section{Preliminaries}

%An \emph{alphabet} is a set of symbols.  A \emph{string} is a sequence of symbols.  A \emph{language} over an alphabet is a set of strings that consist of the symbols from that alphabet.

%A \emph{decision problem} is a template for yes-or-no questions.  An \emph{instance} of a decision problem is specified by a string.  A decision problem is \emph{decidable}, \textiff{} the answer to any instance can be computed given enough, but finite, time and space.  A decision problem is \emph{recognizable}, \textiff{} whenever the answer is ``yes'' to an instance, it can be computed in finite amount of time and space.

%Decision problems are commonly formulated as sets of strings that satisfy a predicate; strings that specify instances for which the answer is ``yes''.

The reader is assumed to be familiar with the standard concepts of automata theory, Turing machines (\TM{}s), and basic complexity classes~\cite{sipser}.

%We will use the terms \emph{algorithm} and the Turing machine interchangeably.  We will call the functions that map the languages to algorithms a \emph{method}.

The following notation will be used throughout this paper:
\begin{itemize}[label=\textbullet]
	\item $\P{A}$ is the power set of $A$.
	\item $A \sqcup B$ is the union of sets $A$ and $B$, that also asserts that the two are disjoint.
	\item % $\sigma\ext\tau$, or simply
	    $\sigma\tau$ is the sequences $\sigma$ and $\tau$ concatenated.
	\item $\sigma_i$ is the \ith{i} element of the sequence $\sigma$.
	\item $\enc{O_1, \dotsc, O_k}$ is the encoding of objects $O_i$ in the alphabet of context.
	% \item $\L{M}$ denotes the language recognized by the machine $M$.
	% \item $\wo{S}{q}$ is the set $S$ without its element $q$
	% \item $\subseq{\sigma}{i}{j}$ is the sub-sequence $(\sigma_i, \dotsc, \sigma_j)$
	% \item $\trim{\sigma}$ is the sequence $\sigma$ without its last element
	% \item % $\sigma \ext x$, or simply
	    % \sigma x$, is the sequence $\sigma$ extended with the element $x$
\end{itemize}

\subsection{Multihead finite automata}

A (two-way) $k$-head nondeterministic finite automaton, denoted \tnfa{k}, is a 6\nobreakdash-tuple consisting of
\begin{enumerate}
    \item a finite set of states $Q$;
    \item an input alphabet $\Sigma$;
    \item a transition function $\delta \colon Q \times \Gamma^k \to \P{Q \times \Delta^k}$, where
    \begin{itemize}[topsep=.5ex,label=\textbullet]
        \item $\Gamma = \Sigma \sqcup \Set{\lend, \rend}$ is the tape alphabet, where $\lend$ and $\rend$ are respectively the left and right end markers, and
        \item $\Delta = \Set{\movel, \movep, \mover}$ is the set of head movements, where \movel{} and \mover{} respectively indicate moving left and right, and \movep{} indicates staying put;
    \end{itemize}
    \item an initial state $q_0 \in Q$;
    \item an accept state $\qacc \in Q$; and
    \item a reject state $\qrej \in Q$.
    % \item a set of final states $F \subseteq Q$.
\end{enumerate}
The \MM{2} in the denotation \tnfa{k} indicates that these automata can move their heads in both directions, i.e.\ that their heads are two-way.  For the rest of the paper, unless specified otherwise, our (multi-head) finite automata should be assumed as two-way.

A \tnfa{k} $M = (Q, \Sigma, \delta, q_0, \qacc, \qrej)$ starts from the state $q_0$ with $\lend x \rend$ written on its single read-only tape where $x \in \Sigma^*$ is the input string.  All $k$ tape heads are initially on the $\lend$ symbol.  The function $\delta$ maps the current state and the $k$ symbols under the tape heads to a set of alternative steps $M$ can take.  By picking an alternative $(q, d)$, $M$ transitions into the state $q$ and moves its \ith{i} head by $d_i$.

% What about attempting moves beyond the end marker????

The \emph{configuration} of a \tnfa{k} $M$ at a step of its execution is the $(k+1)$-tuple consisting of its state and its head  positions at that moment.  The initial configuration of $M$ is $(q_0, 0^k)$. % States $\qacc$ and $\qrej$ are the \emph{halting states} of $M$, and configurations with those are its \emph{halting configurations}. %  A configuration is \emph{accepting} if its state is $\qacc$.

Starting from its initial configuration and following different alternatives offered by $\delta$, a \tnfa{k} $M$ may have several \emph{computational paths} on the same string.  A computational path of $M$ \emph{halts} if it reaches $\qacc$ or $\qrej$, or if $\delta$ does not offer any further steps for $M$ to follow.  $M$ \emph{accepts} an input string $x$ if there is a computational path of $M$ running on $x$ that halts on $\qacc$.  $M$ \emph{rejects} an input string $x$ if $M$ running on $x$ halts on a state other than $\qacc$ on every computational path.  The \emph{language} \emph{recognized by} $M$ is the set of all strings accepted by $M$.

Given an input string $x$, $M$ may have computational paths that never halt.  In the special case that $M$ halts on every computational path for every input string,  $M$ is said to be an \emph{always halting} \tnfa{k}.

A (two-way) $k$-head \emph{deterministic} finite automaton, denoted \tdfa{k}, differs from a \tnfa{k} in its transition function, which is defined as $\delta \colon Q \times \Gamma^k \to Q \times \Delta$.  $1$\nobreakdash-head finite automata are simply called finite automata and are denoted as \dfa{} and \nfa{} for the deterministic and nondeterministic counterparts, respectively.

For any $k$, let \TNFA{k} denote the class of languages recognized by a \tnfa{k}.  
%For when $k$ is unspecified, let \TNFAS{} denote the class of languages that has a \tnfa{k} recognizing them for some $k$.  In other words;
%\begin{equation*}
%    \TNFAS = \bigcup_{k > 0} \TNFA{k}
%\end{equation*}
\TNFA{1} is the class of regular languages~\cite{kutrib_multi}. % SÜSLÜNOTASYON PLZ!!!!!!!!!!!!!!!!!!!!!!!

For any growth function $f(n)$, \NSPACE{f(n)} denotes the class of languages recognized by nondeterministic Turing machines (\NTM{}s) which are allowed to use \OH{f(n)} space for inputs of length $n$.  The class $\NSPACE{\log n}$ is commonly denoted as \NL.

\begin{lemma}\label{lem:nfakisnl}
	Nondeterministic multi-head finite automata are equivalent to logarithmic space \NTM{}s in terms of language recognition power~\cite{hartmanis}.  Put formally;
	\begin{equation*}
%		\TNFAS 
\bigcup_{k > 0} \TNFA{k}= \NL{}.
	\end{equation*}
\end{lemma}

\begin{lemma}\label{lem:nfakhierarchy}
    The languages in \NL{} are organized in a strict hierarchy, based on the number of heads of the nondeterministic finite automata recognizing them~\cite{monien_two-way_1980}.  Formally, the following is true for any $k > 0$:
    \begin{equation*}
        \TNFA{k} \subsetneq \TNFA{k + 1}
    \end{equation*}
\end{lemma}

For any given $k$, let \TNFAT{k}{f(n)} denote the class of languages that are recognized by a \tnfa{k} running for \OH{f(n)} steps on every alternative computational path on any input of length $n$.  Clearly, those machines are also always halting. Let \TNFAST{f(n)} denote the class of languages that are recognized by a nondeterministic multi-head finite automata with any number of heads and running in \OH{f(n)} time.  We use \timel{linear} designation instead of $f(n) = n$.

\begin{lemma}\label{lem:nfakarehalting}
    The following is true for any $k > 0$:
    \begin{equation*}
         \TNFA{k} \subseteq \TNFAT{2k}{n^k}
    \end{equation*}
\end{lemma}

\begin{proof}
    Let $M$ be any \tnfa{k} with $Q$ as its set of states.  Running on an input string of length $n$, $M$ can have $T = \abs{Q} \cdot (n+2)^k$ different configurations.  If $M$ executes  more than $T$ steps, then it must have repeated a configuration.  Therefore, for every input string it accepts, $M$ should have an accepting computational path of at most $T$ steps.
    
    With the help of $k$ additional \emph{counter} heads, the \tnfa{2k} $M'$ can simulate $M$ while imposing it a runtime limit of $T$ steps.  Machine $M'$ can count up to $T$ as follows:  Let $c_1, \dotsc, c_k$ denote the counter heads.  Head $c_1$ moves right every \ith{\abs{Q}} step of $M$'s simulation.  For all $i < k$, whenever the head $c_i$ reaches the right end marker, it rewinds back to the left end, and head $c_{i+1}$ moves once to the right.  If $c_k$ attempts to move past the right end, $M'$ rejects.

	If the simulation halts before timeout, $M'$ reports $M$'s decision.
	The strings that $M$ would loop on are rejected by $M'$ due to timeout.
	The \tnfa{2k} $M'$ recognizes the same language as $M$, but within the time limit of $\OH{n^k}$.
    %\qed
\end{proof}

\Cref{lem:nfakisnl,lem:nfakarehalting} can be combined into the following useful fact.

\begin{corollary}\label{cor:mink}
    For every $A \in \NL$, there is a minimum number $\optk{A}$ such that there exists an always halting \tnfa{\optk{A}} recognizing $A$, but not an always halting \tnfa{h} where $h < \optk{A}$.
\end{corollary}

\begin{proof}
	Let $K_A$ be the set of numbers of heads of always halting multi-head nondeterministic finite automata recognizing $A$.
	By \cref{lem:nfakisnl,lem:nfakarehalting}, for some $k$, there is a \tnfa{k} and thereby an always halting \tnfa{2k} recognizing $A$, respectively.
	Thus, $2k \in K_A$ and $K_A$ is non-empty.
	By the well-ordering principle, $K_A$ has a least element, which we call \optk{A}.
\end{proof}

\begin{lemma}\label{lem:decnfaisdecidable}
    $\DECIDER{\nfa{}} = \Set{ \enc{M} | M \text{ is an always halting \nfa{}}}$ is decidable.
\end{lemma}

\begin{proof}
%[Proof\hspace{0.5ex}\protect\footnotemark\hspace{-0.5ex}]
    The \emph{two-way alternating finite automaton}, denoted \afa{}, is a generalization of the \nfa{} model.  The state set of a \afa{} is partitioned into \emph{universal} and \emph{existential} states.  A \afa{} accepts a string $x$ \textiff{} starting from the initial state, every alternative transition from the universal states and at least one of the alternative transitions from the existential states leads to acceptance.  Thus, a \nfa{} is a \afa{} with only existential states.  We refer the reader to~\cite{lipton_ladner_stock} for a formal definition of the \afa{} model.
    
    A \emph{one-way nondeterministic finite automaton}, denoted \onfa{}, is a \nfa{} that cannot move its head to the left.  A \odfa{} is a deterministic \onfa{}.

    Consider the following algorithm to recognize $\DECIDER{\nfa}$:
    \begin{turing}{D}{On input $\enc{M}$, where $M$ is a \nfa{}, and $\Sigma$ is its alphabet:}
    \item Construct a \afa{}  $M'_{\afa}$ by modifying $M$ to accept whenever it halts and designating every state as universal.
    \item\label{itm:afatoonfa} Convert $M'_{\afa}$ to an equivalent \onfa{} $M'_{\onfa}$.
    \item\label{itm:onfatoodfa} Convert $M'_{\onfa}$ to an equivalent \odfa{} $M'_{\odfa}$.
    \item\label{itm:univdfa} Check whether $M'_{\odfa}$ recognizes $\Sigma^*$.  If it does, \accept.  Otherwise, \reject.
    \end{turing}
    
	By its construction, $M'_{\afa}$ (and therefore $M'_{\odfa}$) recognizes $\Sigma^*$ \textiff{} $M$ halts in every computational path while running on every possible input string, i.e.\ it is always halting.  \Steps{itm:afatoonfa}{itm:onfatoodfa} can be implemented by the algorithms given in~\cite{geffert} and the proof for the Theorem~1.39 of~\cite{sipser}, respectively.  The final check in \step{itm:univdfa}, also known as the universality problem for \odfa{}'s, is decidable in nondeterministic logarithmic space~\cite{kutrib_single}, thus in polynomial time by Corollary~8.26 in~\cite{sipser}.  So the algorithm $D$ decides whether a given \nfa{} $M$ is always halting.
    %\qed
\end{proof}

\subsection{Probabilistic Turing machines and finite automata}\label{subsec:pfa}

A probabilistic Turing machine (\PTM) is a Turing machine equipped  with a randomization device.  In its designated coin-tossing states, a \PTM{} obtains a random bit using the device and proceeds by its value.  The language of a \PTM{} is the set of strings that it accepts with a probability greater than $\sfrac12$.

A (two-way) probabilistic finite automaton (\pfa{}) is a restricted \PTM{} with a single read-only tape.  This model can also be viewed as an extension of a \dfa{} with designated coin-tossing states.%
\footnote{One may also think of a \pfa{} as a \nfa{} where each state has probabilities associated with each of its outgoing transitions, and the machine selects which transition to follow with these corresponding probabilities. To make this alternative model equivalent to the constant-randomness machines studied in this paper, it is sufficient to restrict the transition probabilities to dyadic rationals.}
A \pfa{} tosses a hypothetical coin whenever it is in one of those states and proceeds by its random outcome.  Formally, a \pfa{} consists of the following:
\begin{enumerate}
    \item A finite set of states $Q = \detmark{Q} \sqcup \prmark{Q}$, where
    \begin{itemize}
        \item $\detmark{Q}$ is the set of deterministic states, %\footnotemark{}
			and
        \item $\prmark{Q}$ is the set of coin-tossing states.%\footnotemark[\thefootnote]
        % \footnotetext{The letters \emph{d} and \emph{r} stand for \emph{deterministic} and \emph{random}, respectively.}
    \end{itemize}
    \item An input alphabet $\Sigma$.
    \item A transition function overloaded as deterministic $\detmark\delta$ and coin-tossing $\prmark\delta$, where
    \begin{itemize}
        \item $\detmark\delta \colon \detmark{Q} \times \Gamma \to Q \times \Delta$, where $\Gamma$ and $\Delta$ are as defined for the \tnfa{k}'s, and
        \item $\prmark\delta \colon \prmark{Q} \times \Gamma \times R \to Q \times \Delta$, where $R = \Set{0, 1}$ is a random bit provided by a ``coin toss''.%\unote{This is slightly different in Sipser. There, the coin tossing states have only two possible outcomes.}
    \end{itemize}
    \item An initial state $q_0$.
    \item An accept state $\qacc$.
    \item A reject state $\qrej$.
    % \item a set of final states $F \subseteq Q$.
\end{enumerate}

The language of a \pfa{} is similarly the set of strings which are accepted with a probability greater than $\sfrac12$.

Due to its probabilistic nature, a \PTM{}  may occasionally err and disagree with its language.  In this paper, we will be concerned about the following types of error:
\begin{enumerate}
    \item \emph{Failing to accept} -- rejecting or looping indefinitely given a member input
    \item \emph{Failing to reject} -- accepting or looping indefinitely given a non-member input
\end{enumerate}

\subsection{Interactive proof systems}

Our definitions of interactive proof systems (\IPS{}es) are based on~\cite{dworkstock}.
We will focus on a single variant, namely the private-coin one-way \IPS{} with a finite-state verifier.

An \IPS{} consists of a \emph{verifier} and a \emph{prover}.  The verifier is a \PTM{} vested with the task of recognizing an input string's membership, and the prover is a function providing the purported proof of membership.

In a \emph{private-coin one-way \IPS}, the coin flips (both their outcomes and the information on when they are flipped) are hidden from the prover $P$, and $P$ communicates the proof to the verifier $V$ in a monologue.  In such an \IPS, as a simplification, $P$ can instead be viewed as a certificate function $c \colon \Sigma^* \to \Lambda^\infty$ that maps input strings to infinitely long certificates, where $\Sigma$ and $\Lambda$ are respectively the input and certificate alphabets.   $V$, in turn, can be thought of as having an additional certificate tape to read with a head that cannot move to the left.  Given an input string $x \in \Sigma^*$, $V$ executes on it as usual with $c(x)$ written on its certificate tape.

Note that the ``one-way'' denotation for an \IPS{} qualifies only the interaction (i.e.\ specifies that the verifier does not communicate back), and not the head movements of the verifier.

In this paper, the term ``\PTM{} verifier in a private-coin one-way \IPS{}'' will be abbreviated as ``\PTM{} verifier''.  Accordingly, ``\pfa{} verifier'' shall mean ``two-way probabilistic finite automaton verifier in a private-coin one-way \IPS{}''.

The language $A$ of a \PTM{} verifier $V$ is the set of strings that $V$ can be ``convinced'' to accept with a probability greater than $\sfrac12$ by some certificate function $c$.  The error bound\footnote{Our definition of the error bound corresponds to the ``strong'' version of the \IPS{} definition in~\cite{dworkstock}.} of $V$, denoted $\errf{V}$, is then defined as the minimum value satisfying both of the following:
\begin{itemize}
    \item $\forall x \in A$, $V$ paired with some $c(x)$ accepts $x$ with a probability at least $1 - \errf{V}$.
    \item $\forall x \notin A$, $V$ paired with any $c(x)$ rejects $x$ with a probability at least $1 - \errf{V}$.
\end{itemize}

Let \IPerrl{\err}{t(n),s(n),r(n)} be the class of languages that have verifiers with an error at most $\err$  ($\err < \sfrac12$)  
using \OH{s(n)} space and \OH{r(n)} amount of coins in the worst case and with an expected runtime in \OH{t(n)}, where $n$ denotes the length of the input string.  Instead of a function of $n$, we write simply \lcons, \llog, \lpoly, and \lexp{} to describe constant, logarithmic, polynomial, and exponential limits in terms of the input length, respectively.  We write $0$ and $\infty$ to describe that a resource is unavailable and unlimited, respectively. Furthermore, let
\begin{align*}
    \IPl{t(n), s(n), r(n)} &= \bigcup_{\makebox[\widthof{$_{\err > 0}$}][c]{$_{\err < \frac12}$}} \IPerrl{\err}{t(n), s(n), r(n)},\\
    \shortintertext{and}
    \IPstarl{t(n), s(n), r(n)} &= \bigcap_{\err > 0} \IPerrl{\err}{t(n), s(n), r(n)}.
\end{align*}

% Class \NP{} equals \IPerrl{0}{\lpoly, \infty, 0} by definition.  Since the space usage may not exceed the time limit, we have:
% \begin{equation*}
	% \NP = \IPerrl{0}{\lpoly, \lpoly, 0}
% \end{equation*}

% The following are true by definition:
% \begin{align*}
	% \NP &= \IPl{\lpoly, \infty, 0}\\
	% \NL &= \IPl{\infty, \llog, 0}
% \end{align*}

The following are trivial:
\begin{align*}
	\NP &= \IPl{\lpoly, \lpoly, 0}\\
	\NL &= \IPl{\lpoly, \llog, 0}
	\intertext{The class \NP{} is further characterized~\cite{Con93,cl95} as}
	\NP &= \IPl{\lpoly, \llog, \lpoly} = \IPl{\lpoly, \llog, \llog},\\
	\intertext{and the class \NL{}~\cite{sayyakaryilmaz} as}
	\NL &= \IPl{\infty, \lcons, \lcons}.
\end{align*}

% We note the following known results with their citations over the respective equality signs.
% The first of the equalities in each line is trivial:
% \begin{align*}
	% \NP = \IPl{\lpoly, \lpoly, 0} &\stackrel{\text{\cite{cl95}}}{=} \IPl{\lpoly, \llog, \llog} \stackrel{\text{\cite{Con93}}}{=} \IPl{\lpoly, \llog, \lpoly}\\
	% \NL = \IPl{\lpoly, \llog, 0} &\stackrel{\text{\cite{sayyakaryilmaz}}}{=} \IPl{\infty, \lcons, \lcons}
% \end{align*}

For polynomial-time verifiers with the ability to use at least logarithmic space, the class \IPstarl{t(n), s(n), r(n)} is identical to the corresponding class \IPl{t(n), s(n), r(n)}, since such an amount of memory can be used to time one's own execution and reject computations that exceed the time limit, enabling the verifier to run through several consecutively appended copies of  certificates for the same string and deciding according to the majority of the results of the individual controls.  For constant-space verifiers, this procedure is not possible, and the question of whether \IPstarl{\infty, \lcons, \lcons} equals \IPl{\infty, \lcons, \lcons} is nontrivial, as we will examine in the following sections.

\section{Linear-time 2nfa($k$)'s and verification with small error}

In~\cite{sayyakaryilmaz}, Say and Yakary\i{}lmaz showed that membership in any language in \NL{} may be checked by a \pfa{} verifier using some constant number of random coin tosses.  They also showed  how the \emph{weak} error of the verifier can be made arbitrarily small.\footnote{In contrast to the (strong) error definition we use in this paper, the weak error definition (also by~\cite{dworkstock}) does not regard the verifier looping forever on a non-member input as an error.} We will now describe their approach, which forms the basis of our own work.

The method, which we will name \SYW{}, for producing a constant-randomness \pfa{} verifier given any language $A \in \NL$, takes an always halting \tnfa{k} $M_A$ recognizing $A$ (for some $k$), which exists by \cref{lem:nfakisnl,lem:nfakarehalting}, as its starting point.  The constructed verifier $\SYWf{A}$ will attempt to repeatedly simulate $M_A$ by looking at the certificate while relying on its private coins to compensate for having $k-1$ fewer input heads than $M_A$.  Given any input string $x$, $\SYWf{A}$ expects a certificate $c(x)$ to contain $\C$ successive segments, each of which describe an accepting computational path of $M_A$ on  $x$. $c(x)$ is supposed to provide the following information for each transition of $M_A$ en route to purported acceptance: the symbols read by the $k$ heads, and the nondeterministic branch taken.  Verifier $\SYWf{A}$ attempts to simulate $M_A$ through the provided computational path until either the simulation halts, or $\SYWf{A}$ catches a ``lie'' in the certificate and \emph{rejects}. % To verify $c(x)$ against lies, 
$\SYWf{A}$ chooses a head of $M_A$ at random by tossing $\ceil{\log k}$ coins in private before each simulation.  Throughout the simulation, $\SYWf{A}$ mimics the movements of this chosen head and compares $c(x)$'s claims against what is being scanned by that head, while leaving the remaining $k-1$ unverified.  If the simulation rejects, then so does \SYWf{A}.  If $\C$ such simulation rounds end with acceptance, $\SYWf{A}$  \emph{accepts}.

For any language in $A \in \NL$ which can be recognized by an always halting \tnfa{k} $M_A$, the verifier of \SYW{} simulating $M_A$ for $\C$ rounds tosses a total of $\C \cdot \ceil{\log k}$ coins, which is a constant with respect to the input length.

Paired with the proper certificate $c(x)$, \SYWf{A} accepts all strings $x \in A$ with probability 1.  As mentioned earlier, the ``weak error'' of \SYWf{A} therefore depends only on its worst-case probability of accepting some $x \notin A$.

For $x \notin A$, there does not exist an accepting computation of $M_A$ on $x$.  Still, a certificate may describe a fictional computational path of $M_A$ to acceptance by reporting false values for the symbols read by at least one of the heads.  Since \SYWf{A} cannot check many of the actual readings, it may fail to notice those lies.  However, since \SYWf{A} chooses a head to verify at random, there is a non-zero chance that \SYWf{A} detects any such lie.

The likelihood that \SYWf{A} chooses a head that the certificate is lying about is at least $\sfrac1k$.\footnotemark{}  Therefore, the weak error of \SYWf{A} is at most $\paren*{\sfrac{(k-1)}{k}}^\C$.  This upper bound for weak error can be brought as close to $0$ as one desires by increasing \C, the number of rounds to simulate.

\footnotetext{The error in the approximation $k \approx 2^{\ceil*{\log k}}$ used in this analysis does not affect the end result and simplifies the explanation.}

Although the underlying \tnfa{k} $M_A$ recognizing $A \in \NL$ is an always halting machine, the verifier $\SYWf{A}$ may still be wound up in an infinite loop by some certificate:  $M_A$ might be relying on the joint effort of its many heads to ensure that it always halts.  Since $\SYWf{A}$ validates only a single head's readings, lies on what others read may tamper this joint effort and lead $\SYWf{A}$ into a loop.  A malicious certificate might lead $\SYWf{A}$ in a loop by lying about one head alone.  If this happens during the first round, there would not be any more rounds for $\SYWf{A}$ since it would be in a loop.  The (strong) error $\errf{\SYWf{A}}$ of \SYWf{A} is therefore at most $\sfrac{(k-1)}k$.  This upper bound to $\errf{\SYWf{A}}$ cannot be reduced to less than $\sfrac{(\optk{A}-1)}{\optk{A}}$, where \optk{A} is the minimum number of heads required in an always halting machine to recognize $A$ by \cref{cor:mink}.

Say and Yakary\i{}lmaz also propose the method \SYS{}, which is a slightly modified version of \SYW{} that produces verifiers with errors less than $\sfrac12$, albeit barely so.  Let $A \in \NL$ and $M_A$ be an always halting \tnfa{k} recognizing $A$ for some $k$.  Regardless of the input string, the verifier \SYSf{A} rejects at the very beginning with a probability $\sfrac{(k-1)}{2k}$ by tossing $\ceil{\log{k}} + 1$ coins.  Then it continues just like \SYWf{A}.  The bounds for the error $\errf{\SYSf{A}}$ are as follows:
\begin{equation*}
    \frac{k-1}{2k} \le \errf{\SYSf{A}} \le \frac{k^2 - 1}{2k^2}
\end{equation*}

\subsection{\Kulyutmaz{} and \windable{} heads}

How much of \NL{} may yet fit into \IPstarl{\infty, \lcons, \lcons}?  Method \SYW{} was our starting point in working towards a lower bound for \IPstarl{\infty, \lcons, \lcons}.

Let $M_A$ be the \tnfa{k} that \SYWf{A} uses to verify $A \in \NL$.  The cause for \SYWf{A}'s high strong error turns out to be a decidable characteristic of $M_A$'s heads.  We will refer to such undependable heads as \emph{\windable{}}.

\begin{definition}[\Kulyutmaz{} and \windable{} heads]\label{def:windablehead}
    Let $M$ be a \tnfa{k} with the transition function $\delta \colon Q \times \Gamma^k \to \P{Q \times \Delta^k}$.  
    For $i$ between $1$ and $k$, let $M_i$ be a \nfa{} with the transition function $\delta_i \colon Q \times \Gamma \to \P{Q \times \Delta}$ defined as follows:
    \begin{equation*}
        \delta_i(q, x) = \bigcup_{\subalign{y &\in \Gamma^k\\y_i &= x}}\Set{ (r, d_i) | (r, d) \in \delta(q, y) }
    \end{equation*}
	If $M_i$ is always halting, then the \ith{i} head of $M$ is a \emph{\kulyutmaz{}} head.  Otherwise, it is a \emph{\windable{}} head.
\end{definition}

The execution of each \nfa{} $M_i$ in \cref{def:windablehead} is designed to correspond to the \ith{i}-head-only simulation of the \tnfa{k} $M$ by the verifier of \SYW{}.  Just like the verifier of \SYW{} and by the way $\delta_i$ is defined, $M_i$ can make any of the transitions allowed by $M$'s transition function ($\delta$) and chooses one by the certificate while making sure that the \ith{i} symbol fed to $M$'s transition function ($y_i$) is the same as the symbol it is reading itself ($x$).  Crucially, if a certificate can wind the verifier of \SYW{} into a loop during the single-headed simulation of $M$, then the \nfa{} $M_i$ has a branch of computation that loops with an analogous certificate.  The converse is also true.  Therefore, the verifier of \SYW{} can be wound up in a loop during a round of verification \textiff{} it has chosen a \windable{} head to verify.

\begin{example}
	\def\BendAngle{25}
	\def\BelowDist{1.1cm}
	\def\LoopOut{122}
	\tikzset{
		loop above/.append style={
			every loop/.append style={
				out=\LoopOut, in=180-\LoopOut, looseness=0,
				shorten >= 0pt
			}
		}
	}
	
	Let $\Sigma = \Set{ \texttt0, \texttt1}$ and $\anbn = \Set{ \texttt0^i\texttt1^i | i \in \N }$.  \Cref{fig:manbn} depicts the state diagram of the \tnfa{2} $\manbn = (Q, \Sigma, \delta, q_0, \qacc, \qrej)$:  Whenever $(r, d) \in \delta(q, x)$, an arc is drawn from $q$ to $r$ with the label $x \to d$.  Recall that $x$ and $d$ are tuples of symbols and head movements, respectively.  The arc coming from nowhere to $q_0$ indicates that it is the initial state.  Such a pictorial representation of an automaton is called its \emph{state diagram}.

	\begin{figure}[htbp!]
		\centering
		\begin{tikzpicture}[every edge/.append style={font=\scriptsize}]
			\node[state, initial]	(q0)					{q_0};
			\node[state]			(q1) [right = of q0]	{q_1};
			\node[state]			(q2) [right = of q1]	{q_2};
			\node[state           ]	(qa) [right = of q2]	{\qacc};
			\node[state]			(qr) [below = \BelowDist of qa]	{\qrej};

			\path[->]
				(q0)
				edge			  node [    ] {(\lend, \lend) \to (\mover,\mover)} (q1)
				(q1)
				edge [loop above] node		  {(\texttt0, \texttt0) \to (\movep,\mover)} ()
				edge			  node [    ] {(\texttt0, \texttt1) \to (\mover,\mover)} (q2)
				edge [bend right=\BendAngle] node [swap] {(\rend, \rend) \to (\movep,\movep)} (qa)
				(q2)
				edge [loop above] node		  {(\texttt0, \texttt1) \to (\mover,\mover)} ()
				edge 			  node [    ] {(\texttt1, \rend) \to (\movep,\movep)} (qa);
		\end{tikzpicture}
		\caption{State diagram of \tnfa{2} $\manbn$.}
		\label{fig:manbn}
	\end{figure}
	% \unote{$q_{\text{acc}$'i yuvarlak icine almali miyim?  $q_{\text{rej}$'e ozel bir sey yapmadan sadece $q_{\text{acc}$'e ozel bir sey yapmayi garipsedim.}

	\manbn{} starts by moving its second head to the leftmost \texttt{1} in the input. It then moves both heads to the right as long as they scan  \texttt{0} and \texttt{1}, respectively, and accepts only if this walk ends when these heads simultaneously read \texttt{1} and \rend{}, respectively.  The empty string is accepted immediately.
	% First, it sends its second head to the first \texttt{1} symbol of the input string, while the first head stays put at the beginning.  Then, it moves both of its heads simultaneously to the right, checking if the two heads read \texttt{0} and \texttt{1} respectively; until the first head reads the first \texttt{1} symbol, and the second head reads the \rend{} symbol.  If they do, then \manbn{} accepts.  It also accepts if the string is empty.  In every other case, it rejects due to lack of available transitions.

	\cref{fig:manbn1} depicts \manbn[1], which is the \nfa{} associated with the first head of \manbn{} obtained as described in \cref{def:windablehead}.  It has a computational path that loops indefinitely on $q_1$ given any input string that begins with a \texttt{0}.  Thus, \manbn[1]{} is not always halting, and by \cref{def:windablehead}, the first head of \manbn{} is \windable{}.
	% Is there any input string on which \manbn[1] may loop?  Yes; \manbn[1] indeed has an execution that runs forever on any input string that begins with a \texttt0, via the self-loop at $q_1$.

	\begin{figure}[htbp!]
		\centering
		\begin{tikzpicture}
			\node[state, initial]	(q0)					{q_0};
			\node[state]			(q1) [right = of q0]	{q_1};
			\node[state]			(q2) [right = of q1]	{q_2};
			\node[state           ]	(qa) [right = of q2]	{\qacc};
			\node[state]			(qr) [below = \BelowDist of qa]	{\qrej};

			\path[->]
				(q0)
				edge			  node [    ] {\lend \to \mover} (q1)
				(q1)
				edge [loop above] node		  {\texttt0 \to \movep} ()
				edge			  node [    ] {\texttt0 \to \mover} (q2)
				edge [bend right=\BendAngle] node [swap] {\rend \to \movep} (qa)
				(q2)
				edge [loop above] node		  {\texttt0 \to \mover} ()
				edge 			  node [    ] {\texttt1 \to \movep} (qa);
		\end{tikzpicture}
		\caption{State diagram of \nfa{} \manbn[1].}
		\label{fig:manbn1}
	\end{figure}
	
	Indeed, \SYWf{\anbn} simulating \manbn{} and running on an input string that begins with \texttt0 would loop forever at state $q_1$ if \SYWf{\anbn} were to track the first head, and the certificate were to report infinite sequences of \texttt0's as both heads' readings.

	Unlike the first one, the second head of \manbn{} is \kulyutmaz{}.  \Cref{fig:manbn2} depicts the \nfa{} \manbn[2]{} associated with it.  Since every transition other than those that lead into the accept state moves the head to the right, the head will eventually reach the end of the input, and \manbn[2]{} will accept, unless it gets ``stuck'' (and  implicitly rejects) by encountering a \texttt{0} while at state $q_2$.

	\begin{figure}[htbp!]
		\centering
		\begin{tikzpicture}
			\node[state, initial]	(q0)					{q_0};
			\node[state]			(q1) [right = of q0]	{q_1};
			\node[state]			(q2) [right = of q1]	{q_2};
			\node[state           ]	(qa) [right = of q2]	{\qacc};
			\node[state]			(qr) [below = \BelowDist of qa]	{\qrej};

			\path[->]
				(q0)
				edge			  node [    ] {\lend \to \mover} (q1)
				(q1)
				edge [loop above] node		  {\texttt0 \to \mover} ()
				edge			  node [    ] {\texttt1 \to \mover} (q2)
				edge [bend right=\BendAngle] node [swap] {\rend \to \movep} (qa)
				(q2)
				edge [loop above] node		  {\texttt1 \to \mover} ()
				edge 			  node [    ] {\rend \to \movep} (qa);
		\end{tikzpicture}
		\caption{State diagram of \nfa{} \manbn[2].}
		\label{fig:manbn2}
	\end{figure}

	% For an analogous reason, \SYWf{\anbn} simulating \manbn{} and choosing the second head for its verification also cannot loop, making the second head \kulyutmaz{}.  
\end{example}

\begin{lemma}
    Being \kulyutmaz{} or \windable{} is a decidable property of a \tnfa{k}'s heads.
\end{lemma}

\begin{proof}
    To decide whether the \ith{i} head of a \tnfa{k} $M$ is \kulyutmaz{}, an algorithm can construct the \nfa{} $M_i$ described in \cref{def:windablehead} and test whether $M_i \in \DECIDER{\nfa}$ by the algorithm in \cref{lem:decnfaisdecidable}.
\end{proof}

Consider a language $A \in \NL$ that is recognized by a \tnfa{k} $M_A$ that always halts, and has \kulyutmaz{} heads only.  The verifier \SYWf{A} using $M_A$ cannot choose a \windable{} head, and therefore can never loop.  Thus, it verifies $A$ with $\errf{\SYWf{A}} \le \paren*{\sfrac{(k-1)}{k}}^\C$.

\subsection{\tnfa{k}'s with a \kulyutmaz{} head and small-error verification}

The distinction of \kulyutmaz{} and \windable{} heads has been the key to our improvement to the method \SYW{}.  Method \GB{}, which is to be introduced in the proof of the following lemma, is able to produce verifiers with an error bound equaling any desired non-zero constant for a subset of languages in \NL.

\begin{lemma}\label{lem:singlekulyutmazenough}
    Let $A \in \NL$.  If there exists an always halting \tnfa{k} with at least one \kulyutmaz{} head that recognizes $A$, then $A \in \IPstarl{\infty, \lcons, \lcons}$.
\end{lemma}

\paragraph{Proof idea}
The method \GB{} in the proof will construct verifiers similar to those of \SYW{}, except with a key difference.  Given a language $A \in \NL$ recognized by an always halting \tnfa{k} $M_A$ that has at least one \kulyutmaz{} head, every head of $M_A$ has essentially the same probability of getting chosen by \SYWf{A}.
If $M_A$ does not have any \windable{} heads, then \GBf{A} will be identical to \SYWf{A}.

% Verifier \GBf{A} does the same if $M_A$ does not have any \windable{} heads, in which case looping is impossible, and the error can be reduced arbitrarily simply by increasing the rounds of simulation.

If $M_A$ does have some \windable{} heads, then \GBf{A} will mostly avoid choosing them, although still giving each of them a slight chance to be chosen.
Since \GBf{A} may loop only if it is tracking a \windable{} head, the looping probability of \GBf{A} decreases as the probability of the selected head being \windable{} gets lower.

The redeemable disadvantage of \GBf{A}'s bias towards choosing among \kulyutmaz{} heads is that the certificate's lies about the \windable{} heads are now less likely to be detected.  However, since the bias is not absolute, the probability $p$ of choosing the least likely of the heads is still non-zero.  Thus, the probability of detecting an existing lie in any given round is at least $p$, and \GBf{A}  fails to catch a lie  in $\C$ rounds with probability at most $\paren*{1-p}^\C$, which can be lowered to any non-zero value by increasing $\C$.

It is impossible for \GBf{A} to read multiple rounds of infinite computational paths from the certificate, since \GBf{A} would not be able to get past the first one.  The probability of \GBf{A} looping is the greatest when an infinite computational path is given in the first round.  Thus, the increased number of rounds does not make \GBf{A} any more likely to loop.  

\begin{proof}
	Let $A \in \NL{}$ and $M_A = (Q, \Sigma, \delta, q_0, \qacc, \qrej)$ be an always halting \tnfa{k} recognizing $A$ with at least one \kulyutmaz{} head.
	
	Let $\SB = \ceil{\log k}$.  Let $\kr$ and $\kw$ be the number of \kulyutmaz{} and \windable{} heads, respectively.
	Let $\sffr{i} \in \closedintrange{1}{k}$ be the head index of the \ith{i} \kulyutmaz{} head where $i \in \closedintrange{1}{\kr}$.
	If $\kw > 0$, let $\selfunw$ be defined analogously.

    The following parameters will be controlling the error of the verifier:
    \begin{itemize}[label=\textbullet]
        \item $\C$ as the number of rounds to simulate
        \item $\opw < 1$ as the probability that the selected head is a \windable{} head which must be finitely representable in binary and 0 \textiff{} $\kw$ is zero
    \end{itemize}
    Let $\B$ be the minimum number of fractional digits to represent \opw{} in binary.  Then the algorithm for \GBf{A} is as follows:
	\begin{turing}{\GBf{A}}{On input $x$:}
		\item\label{itm:round}
			Repeat $\C$ times:
			\bitem{Move the tape head to the left end of the input.}
			\bitem{Choose $i$ from \oneto{k} randomly with bias, as follows:}
				\bbitem{Flip $\B$ coins for a uniformly random binary probability value $t$ with $\B$ fractional digits.}
			    \bbitem{Flip $\SB$ more coins for an $\SB$-digit binary number $u < 2^{\SB}$.}
                \bbitem{Let $i = \sffw{(u \bmod \kw) + 1}$ if $t < \opw$, and \sffr{(u \bmod \kr) + 1} otherwise.}
			\bitem{Let $q = q_0$.  Repeat the following until $q = \qacc$:}\label{itm:simend}
				\bbitem{Read $y \in \Sigma^k$ from the certificate.  If $y_i$ differs from the symbol under the tape head, \reject{}.}\label{itm:readingsver}
				\bbitem{Read $(q', d) \in Q \times \Delta^k$ from the certificate.  If $(q', d) \notin \delta(q, y)$, or $q' = \qrej$, \reject{}.}\label{itm:nondetver}
				\bbitem{Set $q = q'$.  Move the tape head by $d_i$.}
		\item\label{itm:accept}
			\Accept.
	\end{turing}
	
	An iteration of \step{itm:round} is called a \emph{round}.  The string of symbols read from the certificate during a round is called a \emph{round of certificate}.  Running on a non-member input string, \GBf{A} \emph{false accepts for a round} when that round ends without rejecting.  Similarly, \GBf{A} \emph{loops on a round} when that round does not end.
	
	Verifier \GBf{A} keeps track of $M_A$'s state, starting from $q_0$ and advancing it by $\delta$ and the reports of the certificate.  At any given round, \GBf{A} can either pass the round by arriving at $\qacc$, \emph{reject} by arriving at $\qrej$ or via verification error, or loop via a loop of transitions availed by $\delta$. % Since the four premises of these three events are disjoint, a certificate may not lead \GBf{A} to any combination of those at the same time, regardless of \GBf{A}'s random choice of head to verify.
	
	Verifier \GBf{A} running on an input string $x \in A$ accepts with probability $1$ when paired with an honest certificate that logs an accepting execution path of $M_A$ for $\C$ rounds.
	
	Given an input $x \notin A$, every execution path of the always halting \tnfa{k} $M_A$ recognizing $A$ rejects eventually.  For \GBf{A} to accept $x$ or loop on it, a certificate $c(x)$ must be reporting an execution path that is possible by $\delta$, despite being impossible for $M$ running on $x$.    The weak point of \GBf{A}'s verification is the fact that it overlooks $k - 1$ symbols in \step{itm:readingsver}.  Hence, $c(x)$ must lie about those overlooked symbols.  Since, however, \GBf{A} chooses a head to verify randomly and in private, any lie about any head has just as much chance of being caught as how often that head gets selected.
	
	Let $p$ be the probability of \GBf{A} choosing the least likely head of $M_A$.  By the restrictions on \opw{} and the definition of \selfunr{} and \selfunw{}, every head of $M_A$ has a non-zero chance of being chosen, and therefore $p > 0$.  If $c(x)$ has a lie in it, then $p$ is also the minimum probability of it being detected.
	
	Falsely accepting a string $x$ is possible for \GBf{A} only if $x$ is not a member of $A$, $c(x)$ lies for more than $\C$ rounds, and \GBf{A} fails to detect the lies in each round.  The probability of this event is at most
	\begin{equation}\label{eq:prfa}
	    \paren*{1-p}^\C.
	\end{equation}
	
	Looping on a string $x \notin A$ is possible for \GBf{A} only if $c(x)$ is a lying certificate with $\C' \le \C$ rounds, \GBf{A} fails to detect the lies in each round, and \GBf{A} chooses a \windable{} head on the final and infinite round.  The probability of this event is at most
	\begin{equation}\label{eq:prloop}
	    \paren*{1-p}^{\C'-1} \cdot \opw.
    \end{equation}
    
    The probability that \GBf{A} falsely accepts (\cref{eq:prfa}) can be reduced arbitrarily to any non-zero value by increasing $\C$.  The probability that it loops on a non-member input (\cref{eq:prloop}) can also be reduced to any positive value by reducing $\opw$ if $\kw > 0$, and is necessarily $0$ otherwise.
    
    Verifier \GBf{A} tosses $m \cdot \paren*{r + s}$ coins; a constant amount that does not depend on the input string.
    %\qed
\end{proof}

In summary, given any language $A \in \NL$ that can be recognized by a \tnfa{k} with at least one \kulyutmaz{} head and for any error bound $\err > 0$, \GBf{A} can verify memberships to $A$ within that bound.  The amount of coins \GBf{A} uses depends on \err{} only and is constant with respect to the input string.

\subsection{Linear-time \tnfa{k}'s and \kulyutmaz{} heads}

\begin{lemma}\label{lem:kulyutmazandlineartime}
	Given a language $A$, the following statements are equivalent:
	\begin{enumerate}[label=(\arabic*)]
		\item\label{itm:havelinearnfak}
			$A \in \TNFAL{k}$.
		\item\label{itm:havekulyutmaz}
			$A$ is recognized by a \tnfa{k} with at least one \kulyutmaz{} head.
			% \item\label{itm:inntispnlogn} $A \in \NTISP{n}{\log n}$.
	\end{enumerate}
\end{lemma}

The proof of \cref{lem:kulyutmazandlineartime} will be in two parts.

\begin{proof}[Proof of \impit{itm:havelinearnfak}{itm:havekulyutmaz}]
	Given $A \in \TNFAL{k}$, for some $k$, there exists a \tnfa{k} $M$ recognizing $A$ together with a constant $c$ such that given any input string $x$, $M$ halts in at most $c \cdot \abs{x}$ steps.  Consider the \tnfa{k + 1} $M'$ which operates its first $k$ heads by $M$'s algorithm and uses its last head $T$ as a timer that moves to the next cell on the input tape every \ith{c} step of the execution.  Head $T$ \emph{times out} when it reaches the end of the string, and $M'$ rejects in that case.

	Note that $M'$ recognizes indeed the same language as $M$ since $M$, as well as $M'$, runs for at most $c \cdot \abs{x}$ steps for any given input string $x$, and therefore $T$ never reaches the end of $x$ nor times out.  Apparent from its monotonic movement, however, head $T$ in $M'$ is \kulyutmaz{}. % because of the explanation that follows.
	% \qed
\end{proof}

\begin{proof}[Proof of \impit{itm:havekulyutmaz}{itm:havelinearnfak}]
	Let $M$ be a \tnfa{k} recognizing $A$ such that its \ith{i} head is a \kulyutmaz{} head.  Let $\delta \colon Q \times \Gamma^k \to \P{Q \times \Delta^k}$ be the transition function of $M$.  Let $M_i$ be the \nfa{} with the following transition function as in  \cref{def:windablehead}:
    \begin{equation*}
        \delta_i(q, x) = \bigcup_{\subalign{y &\in \Gamma^k\\y_i &= x}}\Set{ (r, d_i) | (r, d) \in \delta(q, y) }
    \end{equation*}

    Note the relationship between the computational paths (sequences of configurations) of $M$ and $M_i$ running on the same input string. These machines have the same state set, but $M_i$ is running a program which has been obtained from the program of $M$ by removing all constraints provided by all the other $k-1$ heads.  If one looks at any possible computational path of $M$ through ``filters'' that only show the current state and the present position of the \ith{i} head and hide the rest of the information in $M$'s configurations, one will only see legitimate computational paths of $M_i$. 

    Since the \ith{i} head is \kulyutmaz{}, $M_i$ is always halting, and $\delta_i$ does not allow $M_i$ to ever repeat its configuration in a computation. But this means that   $M$ is also unable to loop forever since the two components of its configuration (the state and the position of its \ith{i} head) can never be in the same combination of values at two different steps. As a result, $M$ cannot run  for more than $\abs{Q}\cdot \paren*{n+2}$ steps, where $n$ is the length of the input string.
	% \qed
\end{proof}

We have proven the following theorem.

\begin{theorem}
    $\TNFASL \subseteq \IPstarl{\infty, \lcons, \lcons}$.
\end{theorem}

Note that the following nonregular languages, among others, have linear-time \tnfa{k}'s and can therefore be verified with arbitrarily small error by constant-randomness, constant-space verifiers:
\begin{align*}
    \anbn &= \Set{ \texttt0^i\texttt1^i | i \in \N }\\
    \PAL &= \Set{ x | x \text{ is the reverse of itself} }\\
    \nanb &= \Set{ x | x \in \Set{ \texttt0, \texttt1 }^* \text{ and contains equally many \texttt0's and \texttt1's} }\\
    \CERT &= \Set{ \TA_1\dotsm\TA_l \tsh \TA_1^+\dotsm\TA_l^+ | l > 0 \text{ and } \TA_1,\dotsc,\TA_l \in \Set{ \texttt0, \texttt1 } }
\end{align*}

There are \tdfa{2}'s without risky heads recognizing the languages \anbn{} and \PAL{}.  We have not been able to find \tnfa{k}'s without risky heads that recognize the languages \nanb{} and \CERT{}.

\section{Towards tighter bounds}

Having determined that $\TNFASL \subseteq \IPstarl{\infty, \lcons, \lcons} \subseteq \NL$, it is natural to ask if any one of these subset relationships can be replaced by equalities. Let us review the evidence we have at hand in this matter.

One approach to prove the claim that constant-space, constant-randomness verifiers can be constructed for every desired positive error bound (i.e.\ that $\IPstarl{\infty, \lcons, \lcons} = \IPl{\infty, \lcons, \lcons}$) would be to show that \NL{} equals \TNFASL, i.e.\ that any \tnfa{k} has a linear-time counterpart recognizing the same language. This, however, is a difficult open question~\cite{sayyakaryilmaz}.  As a matter of fact, there are several examples of famous languages in \NL{}, e.g.
\begin{equation*}
    %\PATH = \Set{ \enc{ G, s, t } | \text{there is a path from node $s$ to node $t$ on the graph $G$} },
    \PATH = \Set{ \enc{ G, s, t } | \text{$G$ is a directed graph with a path from node $s$ to node $t$} },
\end{equation*}
for which we have not been able to construct \tnfa{k}'s with a safe head, and we conjecture that $\TNFASL \neq \NL$.

We will now show that $\TNFASL$ is contained in a subset of \NL{} corresponding to a tighter time restriction of $\OH{\sfrac{n^2}{\log(n)}}$ on the underlying nondeterministic Turing machine. We will use the notation \NTISP{f(n)}{g(n)} for the class of languages that can be verified by a \TM{} that uses \OH{f(n)} time and \OH{g(n)} space, simultaneously.  For motivation, recall that logarithmic-space \TM{}'s require $\Omega\paren*{\sfrac{n^2}{\log(n)}}$ time for recognizing the palindromes language~\cite{cobham,melkebeek,durisgalil}, which is easily recognized by a linear-time \tdfa{2}.

\begin{theorem}\label{thm:linearnfakinquadtimelogspace}
    $\TNFASL \subseteq \NTISP{\sfrac{n^2}{\log(n)}}{\log n}$.
\end{theorem}

\paragraph[Proof idea]{Proof idea}
Given a \tnfa{k} $M$ that runs in linear time, an \NTM{} $N$ can simulate it in \OH{\sfrac{n^2}{\log(n)}} steps.  One such $N$ uses $k$ \emph{\counters} for keeping the head positions of $M$ and $k$ \emph{\caches} for a faster access to the symbols in the vicinity of each head, on a tape with $2k$ tracks.  $N$ initializes its \caches{} with a $\lend$ symbol followed by the first $\log(n)$ symbols of the input and puts a mark on $\lend$ symbols to indicate the position of each simulated head.  \Counters{} are initialized to $0$ for yet another indication of the head positions.

To mimic $M$ reading its tape, $N$ reads the marked symbols on its \caches{}.  To move the simulated heads, $N$ both moves the marks on the \caches{} and adjusts the \counters{}.  If a mark reaches the end of its \cache{}, $N$ \emph{\recaches{}} by copying the $\log(n)$ symbols centered around the corresponding head from the input to that \cache{}.  \Counters{} provide the means for $N$ to locate these symbols on the input.

As the analysis will show, the algorithm described for $N$ runs within the promised time and space bounds.  In the following proof, $N$ will have an additional track that has a mark on its \ith[th]{\sfrac{\log(n)}{2}} cell to indicate the \emph{middle} of the \caches{}.

\begin{proof}
	Let $M = (Q, \Sigma, \delta, q_0, \qacc, \qrej)$ be a \tnfa{k} that runs in linear time.  An \NTM{} $N$ can simulate $M$ by using $2k + 1$ tracks on its tape to have
	\begin{itemize}
	    \item $k$  $\log(n)$-digit binary \counters{}, $\CVAR_1, \dotsc, \CVAR_k$, with their least significant digit on their left end;
	    \item $k$  \caches{} of input excerpts of $\log(n)$ length, $\CACHEVAR_1, \dotsc, \CACHEVAR_k$; and
	    \item a mark on the \ith[th]{\sfrac{\log(n)}{2}} cell to indicate the middle.
	\end{itemize}
	
	The work tape alphabet $\Gamma = \Gamma_{\CVAR}^k \times \Gamma_{\CACHEVAR}^k \times \Set{ \midm, \tblank }$ allows $N$ to encode this information, where
	\begin{itemize}
	    \item $\Gamma_{\CVAR} = \Set{ \texttt0, \texttt1, \tblank }$ to represent each $\CVAR_i$ and
	    \item $\Gamma_{\CACHEVAR} = \tmext\Sigma \sqcup \tmext{\marked\Sigma}$ to represent each \cache, where
	    \begin{itemize}
	        \item $\tmext\Sigma = \Sigma \sqcup \Set{ \lend, \rend, \tsh, \tblank }$ and
	        \item $\tmext{\marked\Sigma}$ is a clone of $\tmext\Sigma$, containing ``marked'' versions of all $\tmext\Sigma$'s symbols.
	    \end{itemize}
	\end{itemize}
	
	Cells of the work tape are initialized with $\tblank^{2k+1}$ symbols.  The algorithm of $N$ is as follows:
	\begin{turing}{N}{On input $x$ of length $n$:}
		\item\label{itm:simnfakinitcounters}
			Write $\texttt0$ to each $\CVAR_i$.
		\item\label{itm:simnfakinitcaches}
            Write $\tsh \marked\lend x_1 \dotsm x_{\log(n)} \tsh$ to each $\CACHEVAR_i$.
		\item\label{itm:simnfakinitmiddle}
            Write \midm{} to the \ith[th]{\sfrac{\log(n)}{2}} cell of the last track.
		\item\label{itm:simnfaksim}
			Let $q = q_0$.  Repeat the following until $q = \qacc$:
			\bitem{Scan the \caches{}. Note the marked symbol in each $\CACHEVAR_i$ as $y_i$ via state transitions.}\label{itm:simnfakread}
		    \bitem{Guess a $(r, d) \in \delta(q, y_1 \dotsm y_k)$.  \Reject{} if the set is empty, or $r = \qrej$.}\label{itm:simnfakguess}
    		\bitem{For all $i$, adjust $\CVAR_i$, and move the mark on $\CACHEVAR_i$ by $d_i$.}\label{itm:simnfakmove}
    		\bitem{\emph{\Recache} each $\CACHEVAR_i$ that has a $\marked\tsh$ symbol as follows:}\label{itm:simnfakrecache}
    		    \bbitem{Clear the mark on $\marked\tsh$ of $\CACHEVAR_i$.}
    		    \bbitem{Go to \ith{\CVAR_i} cell on the input.}\label{itm:simnfaklocateinp}
    		    \bbitem{Go to middle of $\CACHEVAR_i$ on the work tape.}
    		    \bbitem{Move both tape heads left until the left end of $\CACHEVAR_i$ is reached.}
    		    \bbitem{Copy $\log(n)$ symbols from the input to between the $\tsh$ symbols of $\CACHEVAR_i$.}\label{itm:copyrecache}
    		    \bbitem{Move both tape heads left until the middle of $\CACHEVAR_i$ is reached.}
    		    \bbitem{Mark the middle symbol on $\CACHEVAR_i$.}
    		    \bbitem{Set $\CVAR_i$ to the input head's position index.}\label{itm:simnfakcountersetback}
    		\bitem{Update $q$ as $r$.}\label{itm:simnfakstep}
    	\item
    	    \Accept.
	\end{turing}
	
	$N$ should carefully prepend/append the left/right end marker to a cache when copying the beginning/end of the input in \step{itm:copyrecache}, respectively.  $N$ should also skip \step{itm:simnfakmove} for an $i$ if the corresponding movement is done while reading an end marker and attempting a movement beyond it.  These details have been omitted from the algorithm to reduce clutter.
	
	Counting up to $n$ in binary is a common task across this algorithm, and it takes linear time by a standard result of amortized analysis.  Only the stages that take a constant number of steps are omitted from the following analysis.
	
	\Step{itm:simnfakinitcaches} takes \OH{n} time as it involves counting up to $n$ in binary to find and mark the \ith{\log(n)} cell on the \caches{}.  After putting \tsh{} on both ends, copying $x_1\dotsm x_{\log(n)}$ in between them takes $\log(n)$ more steps.  \Step{itm:simnfakinitmiddle} can be performed in \OH{\log^2 n} steps by putting $\midm$ symbols to both ends (aligned with the $\tsh$ symbols) and moving them towards the center one by one until they meet.
	
	Given that $M$ runs in linear time, the loop of \step{itm:simnfaksim} is repeated for at most \OH{n} many times.  \Steps{itm:simnfakread}{itm:simnfakmove} take  logarithmic time.
	
	The \recaching{} in \step{itm:simnfakrecache} is to shift the window of input on a cache by $\sfrac{\log(n)}{2}$, so that the mark will be centered on that cache.  \Steps{itm:simnfaklocateinp}{itm:simnfakcountersetback} are the most time consuming sub-stages of a \recache{}, involving decrementing of $\CVAR_i$ down to 1 and setting it back to its original value, respectively.  They both take \OH{n} time since they count down from or up to $n$ at most.  Every other sub-stage of a \recache{} takes \OH{\log n} time.  As a result, each \recache{} takes \OH{n} time.
	
	\Recaches{} are prohibitively slow.  Luckily, since the head marker moves to the middle with every \recache{}, a subsequent \recache{} cannot happen on the same \cache{} for at least another $\sfrac{\log(n)}{2}$ steps of the simulation.  Moreover, since the number of steps that $M$ runs is in \OH{n}, the number of times a \cache{} can be \recached{} is in $\OH{\sfrac{n}{\log(n)}}$ for the entire simulation.  Hence, \step{itm:simnfakrecache}'s time cost to $N$ is $\OH{\sfrac{n^2}{\log(n)}}$.
	
	\Caches{} and \counters{} occupy \OH{\log n} cells on $N$'s tape.  Since every stage of $N$ runs in \OH{\sfrac{n^2}{\log(n)}} time, so does $N$.
	%\qed
\end{proof}

It is not known whether \NL{} contains any language that is not a member of $\NTISP{\sfrac{n^2}{\log(n)}}{\log n}$.

If $\IPstarl{\infty, \lcons, \lcons}$   is indeed a proper subset of $\IPl{\infty, \lcons, \lcons}$, studying the effects of imposing an additional time-related bound on the verifier may be worthwhile in the search for a characterization. We conclude this section by noting the following relationship between runtime, the amount of randomness used, and the probability of being fooled by a certificate to run forever in our setup:

\begin{lemma}\label{lem:strongerrbound}
    Let $V$ be a \pfa{} verifier that flips at most $r$ coins  in a private-coin one-way \IPS{} for the language $A$.  If some string $x \notin A$ of length $n$ can be paired with some certificate $c(x)$ that causes $V$ to  run for $\omega\paren[\big]{n^{2^{r-1}}}$ steps with probability $1$, then $V$ has error at least $\sfrac12$.
\end{lemma}

\begin{proof}
    Let $V$ be a \pfa{} as described above.  By an idea introduced in~\cite{sayyakaryilmaz}, we will construct a verifier equivalent to $V$.  For $z \in \bin^r$, let $V_z$ be the \dfa{} verifier that is based on $V$, but hard-wired to assume that its \ith{i} ``coin flip'' has the outcome $z_i$.  Construct a \pfa{} verifier $V'$ that flips $r$ coins at the beginning of its execution and obtains the $r$-bit random string $z$.  $V'$ then passes control to $V_z$.
    
    Verifiers $V$ and $V'$ have the same behavior whenever their random bits are the same.  Therefore, they are equivalent.
    
    Each $V_z$ has \THETA{n} different configurations, where $n$ denotes the length of the input string.  Similarly, any collection of $2^{r-1}$ distinct $V_z$ has \THETAp{\big}{n^{2^{r-1}}} different collective configurations.  Let $\mathcal{V}$ be any one of those collections.
    
    Let $x$ and $c(x)$ be a nonmember string and its certificate satisfying the premise of the statement.  Then each $V_z$ paired with $c(x)$ also runs on $x$ for $\omega\paren[\big]{n^{2^{r-1}}}$ steps.  The collection $\mathcal{V}$, in that many steps, necessarily repeats a collective configuration.
    
    Consider the prefix $p(x)$ of $c(x)$ consumed by $V'$ until the first time a collective configuration of $\mathcal{V}$ is repeated.  Also consider the suffix $s(x)$ of $p(x)$ consumed by $V'$ since the first occurrence of the repeated collective configuration.  Then $V'$ paired with the certificate $c'(x) = p(x)s(x)^\infty$ repeats its configurations forever whenever it chooses any of the $V_z \in \mathcal{V}$ to pass the execution to.
    
    Both $V'$ and $V$ paired with $c'(x)$ loop on $x$ with a probability at least $\sfrac12$.  Consequently, their errors are at least $\sfrac12$.
    %\qed
\end{proof}

\section{Open questions}

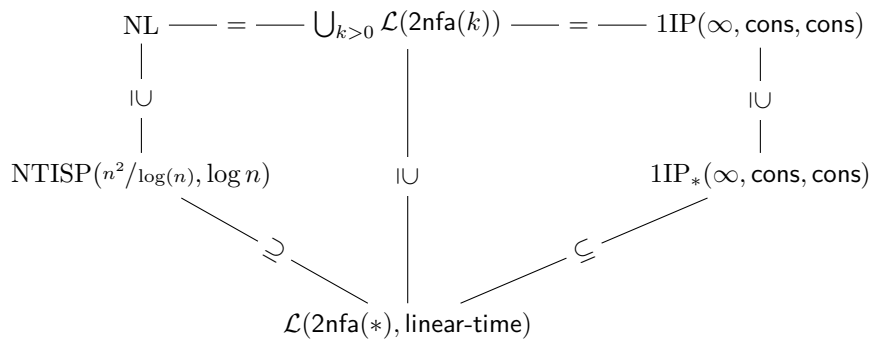
\begin{figure}[htbp!]
	\centering
	\begin{tikzpicture}[every edge quotes/.style={fill=white,sloped,auto=false},
						eqfn/.style={above=-.6em},
						sbfn/.style={above=-.7em},
						node distance = 2cm and 3.5cm]
						%[every node/.style={transform shape}, scale=0.5, initial where=above]
		\node (nl) {\NL};
		\node (tnfas) [right = 1.8cm of nl.east, anchor=west] {\TNFAS};
		\node (ipl) [right = 1.8cm of tnfas.east, anchor=west] {\IPl{\infty, \lcons, \lcons}};
		\node (ipstarl) [below = of ipl] {\IPstarl{\infty, \lcons, \lcons}};
		\node (tnfasl) [below = 4cm of tnfas] {\TNFASL};

		\path%[->]
			(nl) edge ["="' eqfn] (tnfas)
			(tnfas) edge ["="' eqfn] (ipl)
					edge ["\supseteq"] (tnfasl)
			(ipl) edge ["\supseteq"] (ipstarl)
			(tnfasl) edge ["\subseteq"] (ipstarl);

			\node (ntisp) [below = of nl] {\NTISP{\sfrac{n^2}{\log(n)}}{\log n}};
			\path
				(nl) edge ["\supseteq"] (ntisp)
				(ntisp) edge ["\supseteq"] (tnfasl);
	\end{tikzpicture}
	\caption{Inclusion diagram of the language classes covered.}
	\label{fig:inclusiondiagram}
\end{figure}

For an overview of our results, we present the inclusion diagram in \cref{fig:inclusiondiagram}.  The equalities on the left- and right-hand sides of the diagram were shown in~\cite{hartmanis} and~\cite{sayyakaryilmaz}, respectively.  We conclude with a list of open questions.

\begin{itemize}
	\item
		Is there a language in \NL{}, or even in $\NTISP{\sfrac{n^2}{\log(n)}}{\log n}$, requiring any \tnfa{k} recognizing it to have a super-linear runtime?
	\item
		Is there a language in \NL{} that cannot be recognized by any log-space \NTM{} running in $\OH{\sfrac{n^2}{\log(n)}}$ time?
		% Can every log-space \NTM{} be converted into one that runs in $\OH{\sfrac{n^2}{\log(n)}}$ time and recognizes the same language?
	\item
		Is there a language verified by some constant-space, constant-randomness machine, but not by one with smaller strong error? Is it possible to build such a verifier for any language in \NL{} and for any desired positive error bound?
	\item
		Is it possible to construct a linear-time \tnfa{k} for every language that has verifiers using constant space and randomness for any desired positive  strong error?
	% \item
		% Is there a log-space \NTM{} running in $\oh{\sfrac{n^2}{\log(n)}}$ for every language recognized by a \tnfa{k} running in linear time?
\end{itemize}

% To summarize our open questions, we are curious to know whether the improper inclusions in the diagram are equalities or proper inclusions.

\section*{Acknowledgments}
We thank Neal E. Young for the algorithm in the proof of \cref{lem:decnfaisdecidable}. We are grateful to Martin Kutrib for providing us with an outline of the proof of \cref{thm:linearnfakinquadtimelogspace}. We also thank Ryan O'Donnell and Ryan Williams for their helpful answers to our questions, and the anonymous referees for their constructive comments.

\begingroup
\raggedright
\bibliographystyle{elsarticle-num}
\bibliography{ref} 

\begin{thebibliography}{10}
\expandafter\ifx\csname url\endcsname\relax
  \def\url#1{\texttt{#1}}\fi
\expandafter\ifx\csname urlprefix\endcsname\relax\def\urlprefix{URL }\fi
\expandafter\ifx\csname href\endcsname\relax
  \def\href#1#2{#2} \def\path#1{#1}\fi

\bibitem{gezer}
M.~U. Gezer, Windable heads and recognizing {NL} with constant randomness, in:
  A.~Leporati, C.~Mart{\'i}n-Vide, D.~Shapira, C.~Zandron (Eds.), Language and
  Automata Theory and Applications, Springer International Publishing, Cham,
  2020, pp. 184--195.

\bibitem{cl95}
A.~Condon, R.~Ladner, Interactive proof systems with polynomially bounded
  strategies, Journal of Computer and System Sciences 50~(3) (1995) 506--518.

\bibitem{condonlipton}
A.~Condon, R.~J. Lipton, On the complexity of space bounded interactive proofs,
  in: 30th Annual Symposium on Foundations of Computer Science, 1989, pp.
  462--467.

\bibitem{dworkstock}
C.~Dwork, L.~Stockmeyer, Finite state verifiers {I}: The power of interaction,
  J. ACM 39~(4) (1992) 800--828.

\bibitem{nishimura2009}
H.~Nishimura, T.~Yamakami, An application of quantum finite automata to
  interactive proof systems, Journal of Computer and System Sciences 75~(4)
  (2009) 255--269.

\bibitem{nishimura2015}
H.~Nishimura, T.~Yamakami, Interactive proofs with quantum finite automata,
  Theoretical Computer Science 568 (2015) 1--18.

\bibitem{yakaryilmazqAM}
A.~Yakary{\i}lmaz, Public qubits versus private coins, in: Workshop on Quantum
  and Classical Complexity, University of Latvia Press, Riga, 2013, pp. 45--60,
  {ECCC:TR12-130}.

\bibitem{zheng}
S.~Zheng, D.~Qiu, J.~Gruska, Power of the interactive proof systems with
  verifiers modeled by semi-quantum two-way finite automata, Information and
  Computation 241 (2015) 197--214.

\bibitem{feige}
U.~Feige, A.~Shamir, Multi-oracle interactive protocols with constant space
  verifiers, Journal of Computer and System Sciences 44~(2) (1992) 259--271.

\bibitem{debate}
H.~G. Demirci, A.~C.~C. Say, A.~Yakary\i{}lmaz, The complexity of debate
  checking, Theory of Computing Systems 57~(1) (2015) 36--80.

\bibitem{transparent}
A.~Yakary\i{}lmaz, A.~C.~C. Say, H.~G. Demirci, Debates with small transparent
  quantum verifiers, International Journal of Foundations of Computer Science
  27~(02) (2016) 283--300.

\bibitem{sayyakaryilmaz}
A.~C.~C. Say, A.~Yakary\i{}lmaz, Finite state verifiers with constant
  randomness, Logical Methods in Computer Science 10~(3) (Aug. 2014).

\bibitem{sipser}
M.~Sipser, Introduction to the Theory of Computation, Cengage Learning, 2012.

\bibitem{kutrib_multi}
M.~Holzer, M.~Kutrib, A.~Malcher, Complexity of multi-head finite automata:
  {Origins} and directions, Theoretical Computer Science 412~(1-2) (2011)
  83--96.

\bibitem{hartmanis}
J.~Hartmanis, On non-determinancy in simple computing devices, Acta Informatica
  1~(4) (1972) 336--344.

\bibitem{monien_two-way_1980}
B.~Monien, Two-way multihead automata over a one-letter alphabet, RAIRO.
  Inform. th\'eor. 14~(1) (1980) 67--82.

\bibitem{lipton_ladner_stock}
R.~E. Ladner, R.~J. Lipton, L.~J. Stockmeyer, Alternating pushdown automata,
  in: Proceedings of 19th Annual IEEE Symposium on Foundations of Computer
  Science, IEEE Computer Society, 1978, pp. 92--106.

\bibitem{geffert}
V.~Geffert, A.~Okhotin, Transforming two-way alternating finite automata to
  one-way nondeterministic automata, in: International Symposium on
  Mathematical Foundations of Computer Science, Springer, 2014, pp. 291--302.

\bibitem{kutrib_single}
M.~Holzer, M.~Kutrib, Descriptional and computational complexity of finite
  automata---{A} survey, Information and Computation 209~(3) (2011) 456--470.

\bibitem{Con93}
A.~Condon, The complexity of the max word problem and the power of one-way
  interactive proof systems, computational complexity 3~(3) (1993) 292--305.

\bibitem{cobham}
A.~Cobham, Time and memory capacity bounds for machines which recognize squares
  or palindromes, IBM Res. Rep. RC-1621 (1966).

\bibitem{melkebeek}
D.~van Melkebeek, Time-space lower bounds for {NP}-complete problems, in:
  G.~Plun, G.~Rozenberg, A.~Salomaa (Eds.), Current Trends in Theoretical
  Computer Science, World Scientific, 2004, pp. 265--291.

\bibitem{durisgalil}
P.~D{\'u}ri{\'s}, Z.~Galil, A time-space tradeoff for language recognition,
  Mathematical systems theory 17~(1) (1984) 3--12.

\end{thebibliography}
\endgroup

\end{document}